\documentclass[conference,9pt,letterpaper]{IEEEtran}

\newif\ifanonymous
\anonymousfalse

\usepackage{cite}
\usepackage{amsmath,amssymb,amsfonts}
\usepackage{algorithmic}
\usepackage{algorithm}
\usepackage{multirow}
\usepackage{graphicx}
\usepackage{svg}
\usepackage{textcomp}
\usepackage{xcolor}
\usepackage{booktabs}
\usepackage{pifont}

\newcommand{\Pw}{P^{\mathrm{wm}}}

\def\BibTeX{{\rm B\kern-.05em{\sc i\kern-.025em b}\kern-.08em
    T\kern-.1667em\lower.7ex\hbox{E}\kern-.125emX}}
\begin{document}

\title{\fontsize{19}{24}\selectfont HDMark: Holographic Multi-Bit Model Watermarking for Intellectual Property Protection in Hyperdimensional Computing}

\ifanonymous
  \author{\IEEEauthorblockN{Anonymous Authors}
         \IEEEauthorblockA{Paper under double-blind review}}
\else

  \author{
    Hamed Poursiami$^{*}$ \quad
    Ahmad Tahmasivand$^{*}$ \quad
    Farshad Soleiman$^{*}$ \quad
    Guojing Cong$^{\dagger}$ \\
    Thomas Potok$^{\dagger}$ \quad
    Kai Zeng$^{*}$ \quad
    Khaled N. Khasawneh$^{*}$ \quad
    Maryam Parsa$^{*}$ \\[0.5em]
    $^{1}$Electrical and Computer Engineering Department, George Mason University, Fairfax, VA, USA \\
    $^{2}$Oak Ridge National Laboratory, Oak Ridge, TN, USA \\[0.3em]
    {\small\tt{hpoursia@gmu.edu, atahmasi@gmu.edu, fsoleima@gmu.edu, congg@ornl.gov,}}\\
    {\small\tt{potokte@ornl.gov, kzeng2@gmu.edu, kkhasawn@gmu.edu, mparsa@gmu.edu}}
    }
\fi

\maketitle

\begin{abstract}
Hyperdimensional computing (HDC) models deployed on edge hardware are valuable intellectual property that is easy to steal, since a trained classifier consolidates its learned knowledge into a compact set of class prototype hypervectors; yet protecting their ownership through watermarking remains largely unexplored. We present HDMark, the first model watermarking framework native to HDC, which encodes ownership in the same hyperdimensional algebra the model computes with. HDMark maps a multi-bit owner identity to secret quasi-orthogonal carriers, superimposes them holographically onto the trained class prototypes, and verifies ownership with a statistical test. Across four datasets and three HDC learning rules, HDMark leaves accuracy essentially unchanged and verifies a 64-bit identity in $99.8\%$ of trials at $p < 10^{-3}$. We derive closed-form expressions for prediction fidelity and bit recovery that match measurements within $0.4$\,pp mean absolute error. The watermark survives fine-tuning and low-bit quantization, while attempting to erase it through noise injection, pruning, or overwriting severely degrades model accuracy. We further identify a dimension-permutation attack specific to HDC, which reorders hypervector coordinates to evade verification without altering predictions, and counter it through encoder realignment. Finally, an AMD ZCU104 FPGA implementation confirms that HDMark adds no hardware overhead, remains bit-exact under aggressive undervolting, and stays verifiable under timing faults until the classifier itself fails.
\end{abstract}


\section{Introduction}
\label{sec:intro}

Hyperdimensional computing (HDC) has emerged as a lightweight learning paradigm for resource-constrained edge intelligence~\cite{ge2020classification,snyder2026qfhrr,khaleghi2022generic}. By encoding inputs into holographic, high-dimensional representations (\emph{hypervectors}) through simple algebraic operations, HDC enables rapid training, strong tolerance to noise and hardware errors, and highly efficient implementations on FPGAs, ASICs, and in-memory computing platforms~\cite{salamat2019f5,liang2026conformalhdc,poursiami2025scalable,karunaratne2020memory}. Yet this very simplicity makes deployed HDC models easy targets for intellectual property (IP) theft. An HDC classifier consolidates the entire result of proprietary data collection, encoder design, and training into an encoder and a handful of class prototype hypervectors stored in on-chip memory. An adversary with physical or firmware access to an edge device can read them out and replicate the classifier verbatim~\cite{duan2022hdlock}.

Despite this exposure, protecting HDC models as IP remains largely unexplored, and the few existing studies address different problems. StealHD~\cite{ma2026stealing} shows that an HDC model can be extracted through prediction queries without any data, HDLock~\cite{duan2022hdlock} locks the encoder with a key to hinder reverse engineering, and EP-HDC~\cite{park2025ephdc} encrypts the model parameters for privacy-preserving inference. These works demonstrate the threat or restrict access to the model, but none can establish who owns a copy once it has leaked. Model watermarking provides this evidence by embedding a secret signature that the owner can later detect in a suspect model~\cite{uchida2017embedding,poursiami2024watermarking,soleiman2024randomized,adi2018turning}. The only study of watermarking in HDC, a concurrent work~\cite{poursiami2026intellectual}, transfers the trigger-set approach of DNNs, in which the model is trained to assign secret labels to secret inputs. This allows query-only verification, but the watermark must be learned together with the task, which alters the model's predictions and is unreliable under single-pass HDC learning, and it indicates only that a mark is present without carrying an owner identity.

To address these limitations, we present \emph{HDMark}, a multi-bit watermarking framework specifically tailored for HDC classifiers that operates directly in the hyperdimensional algebra of the model. HDMark hashes the owner's identity to a multi-bit payload and embeds it holographically into the trained class prototypes using secret quasi-orthogonal carriers, without retraining. Verification requires only the owner's key and the prototypes of the suspect model, allowing the payload to be recovered by projecting the prototypes onto the carriers and evaluated with a statistical test. Our contributions are:

\begin{itemize}
  \item \textbf{Native HDC model watermarking:} We introduce HDMark, the
  first native, multi-bit model watermarking framework for HDC, applied
  post-training without retraining. HDMark embeds a 64-bit owner
  identity with at most $0.05$\,pp accuracy change, at least $99.6\%$
  prediction agreement, and $99.8\%$ average verification at
  $p < 10^{-3}$.

  \item \textbf{Closed-form analysis:} We derive closed-form expressions
  for prediction fidelity and bit recovery rate from the hyperdimensional
  geometry of the watermarked prototypes, matching empirical
  measurements within $0.4$\,pp mean absolute error.

  \item \textbf{Attack resilience:} We show that fine-tuning, noise
  injection, pruning, overwriting, and quantization cannot remove the
  watermark without substantially degrading model accuracy. We further
  identify a function-preserving dimension-permutation attack specific
  to HDC and counter it via encoder realignment.

  \item \textbf{FPGA implementation and fault evaluation:} We implement
  inference and verification on an AMD ZCU104 FPGA, reusing the
  inference datapath with no extra hardware and no measurable energy
  overhead. HDMark remains bit-exact across the full undervolting range
  down to $0.652$\,V, and under timing faults it preserves verification
  until the classifier degrades or the hardware hangs, with zero false
  claims.
\end{itemize}

\section{Preliminaries: HDC}
\label{sec:prelim}

Hyperdimensional computing (HDC) is a brain-inspired computing paradigm that represents information as high-dimensional vectors, called \emph{hypervectors}, of dimension $D$ (typically $10^3-10^4$)~\cite{kanerva2009,kleyko2022survey}. The representation is \emph{holographic} (i.e., information is distributed across all $D$ components), and relies on the fact that independently drawn random hypervectors are \emph{quasi-orthogonal}, with cosine similarity concentrating around zero with standard deviation $1/\sqrt{D}$. HDC composes and manipulates such representations with three lightweight operations. \emph{Bundling} (e.g., element-wise addition) superimposes hypervectors into one that remains similar to each constituent; a bundle of $n \ll D$ quasi-orthogonal unit hypervectors has similarity of about $1/\sqrt{n}$ to every one of them. \emph{Binding} (e.g., element-wise multiplication, $\odot$) associates two hypervectors into one that is dissimilar to both. \emph{Permutation} (typically a cyclic shift) encodes order or position.

\textbf{Encoding.} An HDC classifier first maps an input
$x \in \mathbb{R}^{F}$ to a hypervector $h = \phi(x) \in \mathbb{R}^{D}$. Two general-purpose encoders for feature vectors are representative. Record-based encoding~\cite{imani2017voicehd} binds a random bipolar key $k_j$ of each feature to a level hypervector $\ell(x_j)$ of its quantized value and bundles the result, $h = \sum_{j=1}^{F} k_j \odot \ell(x_j)$, optionally followed by $\mathrm{sign}(\cdot)$ to keep $h$ bipolar. Nonlinear random projection computes $h = \cos(Wx + \beta) \odot \sin(Wx)$ with Gaussian $W \in \mathbb{R}^{D \times F}$ and $\beta \sim \mathrm{Uniform}[0, 2\pi)^{D}$~\cite{zou2021scalable}. Specialized encoders exist for structured data such as graphs, text, and biosignals~\cite{poduval2022graphd,cong2026hypergraphx,poursiami2026vsgraph,joshi2016language,rahimi2016biosignal}.

\textbf{Inference.} The model stores one prototype
$P_c \in \mathbb{R}^{D}$ per class $c \in \{1,\dots,C\}$ and predicts the
class whose prototype is most similar to the encoded input:
\begin{equation}
\hat{y}(x) = \arg\max_{c}\; \delta\big(\phi(x), P_c\big), \qquad
\delta(u,v) = \frac{\langle u, v\rangle}{\lVert u\rVert\,\lVert v\rVert}.
\label{eq:inference}
\end{equation}

\textbf{Prototype learning.} Given training pairs $(x_i, y_i)$ with $h_i = \phi(x_i)$, three representative learning rules span single-pass aggregation to iterative, error-driven refinement. \emph{Vanilla HDC}~\cite{kanerva2009,rahimi2016robust} bundles the encoded samples of each class in a single pass,
\begin{equation}
P_c = \sum_{i:\, y_i = c} h_i .
\label{eq:vanilla}
\end{equation}
\emph{AdaptHD}~\cite{imani2019adapthd} and \emph{OnlineHD}~\cite{hernandez2021onlinehd} iterate over the training set for several epochs and update the prototypes only when a sample is mispredicted ($\hat{y}_i \neq y_i$), reinforcing the true class and penalizing the predicted one:
\begin{equation}
P_{y_i} \leftarrow P_{y_i} + \lambda_{y_i} h_i, \qquad
P_{\hat{y}_i} \leftarrow P_{\hat{y}_i} - \lambda_{\hat{y}_i} h_i .
\label{eq:update}
\end{equation}
AdaptHD uses a constant step $\lambda_{y_i} = \lambda_{\hat{y}_i} = \eta$, whereas OnlineHD scales each step by how dissimilar the sample is to the prototype, $\lambda_{c} = \eta\,\big(1 - \delta(h_i, P_c)\big)$.

\section{HDMark}
\label{sec:method}

\subsection{Threat Model and Overview}
We consider an owner who trains an HDC classifier and deploys it on devices outside their control. An adversary obtains a full copy of the model, i.e., the encoder and the prototypes, for example by reading the device memory. The adversary does not know the owner's secret key and wants to keep the model useful, but may alter the prototypes to destroy ownership evidence: fine-tuning with limited data, pruning, quantization, noise injection, embedding a watermark of their own, or permuting the hypervector dimensions. In a dispute, the owner or an arbiter can read the prototypes of the suspect model.

\begin{figure*}[htbp]
  \centering
  \includegraphics[width=\linewidth]{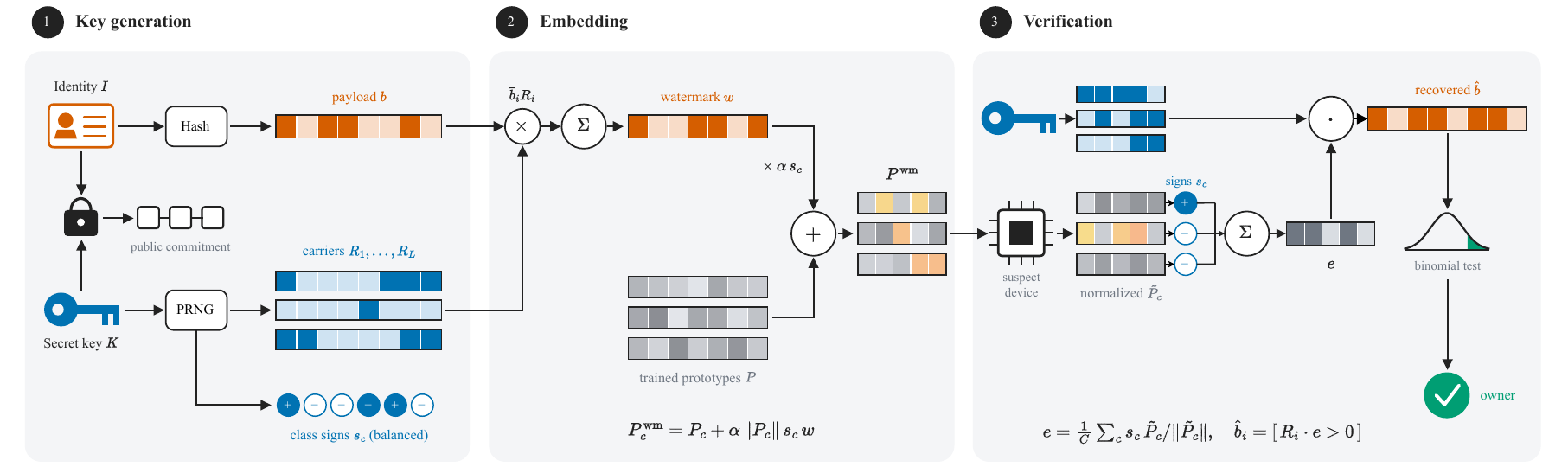}
  \caption{Overview of HDMark. \ding{182}~Key generation: the identity $I$ is hashed to the payload $b$, and the secret key $K$ seeds the carriers $R_i$ and the balanced class signs $s$. \ding{183}~Embedding: the carriers, signed by the payload bits, are bundled into the watermark $w$ and superimposed on the trained prototypes. \ding{184}~Verification: the payload is recovered from the suspect prototypes using the key and claimed identity, and tested with a binomial test.}
  \label{fig:method}
\end{figure*}

HDMark has three phases (Fig.~\ref{fig:method}): key generation, embedding, and verification. It aims at \emph{fidelity} (predictions are preserved), \emph{reliability} (the owner's payload is recovered), \emph{integrity} (a model that does not carry the watermark is not claimed), and \emph{robustness} (the watermark persists as long as the model remains useful).

\subsection{Key Generation}
The owner forms an identity string $I$ (e.g., owner, licensee, and model identifiers) and hashes it to an $L$-bit payload $b \in \{0,1\}^{L}$ (Fig.~\ref{fig:method}, \ding{182}). Hashing binds the payload to the identity, and distinct identities, e.g., one per licensee, yield distinct payloads. From a secret key $K$, a pseudorandom generator produces $L$ bipolar \emph{carriers} $R_i \in \{\pm 1/\sqrt{D}\}^{D}$ and \emph{class signs} $s \in \{\pm 1\}^{C}$ that are balanced, $\sum_{c} s_c = 0$ (for odd $C$, the two groups differ by one). Being random hypervectors, the carriers are quasi-orthogonal to one another and to any hypervector of the model, so the $L$ bits can share the same $D$ dimensions with little mutual interference.
To prevent post-hoc fabrication of an ownership claim, in which an adversary searches for a key and identity that happen to pass verification on a model they do not own, the owner commits to $(K, I)$ before releasing the model, e.g., by timestamping a cryptographic hash of them on a public ledger or with a trusted authority~\cite{adi2018turning}.

\subsection{Embedding}
Each payload bit is mapped to bipolar form, $\bar{b}_i = 2b_i - 1$, and
multiplied by its carrier; the results are bundled and normalized into
a single watermark hypervector (Fig.~\ref{fig:method}, \ding{183}):
\begin{equation}
w = \frac{\sum_{i=1}^{L} \bar{b}_i R_i}{\big\lVert \sum_{i=1}^{L} \bar{b}_i R_i \big\rVert}
\;\approx\; \frac{1}{\sqrt{L}} \sum_{i=1}^{L} \bar{b}_i R_i ,
\label{eq:w}
\end{equation}
where the approximation holds because the carriers are quasi-orthogonal.
The watermark is then superimposed on every prototype:
\begin{equation}
\Pw_c = P_c + \alpha \|P_c\| s_c w, \qquad c = 1,\dots,C .
\label{eq:embed}
\end{equation}
This is a single bundling step applied after training; it needs no
training data and no retraining.

\emph{Relative amplitude.} Inference in Eq.~\eqref{eq:inference} depends
only on the direction of each prototype, and prototype norms differ
widely across classes and learning rules. Setting the amplitude relative
to $\lVert P_c \rVert$ makes $\alpha$ the same relative perturbation for
every prototype.

\emph{Fidelity.} For an encoded input $h$, Eq.~\eqref{eq:embed} gives
\begin{equation}
\delta(h, \Pw_c) = \frac{\delta(h, P_c) + \alpha\, s_c\, \delta(h, w)}
{\sqrt{1 + \alpha^2 + 2\alpha\, s_c\, \delta(P_c, w)}} .
\label{eq:fidelity}
\end{equation}
Because $w$ is quasi-orthogonal to both $h$ and $P_c$, the terms
$\delta(h,w)$ and $\delta(P_c,w)$ are of order $1/\sqrt{D}$. Every class
similarity is therefore scaled by nearly the same factor
$1/\sqrt{1+\alpha^2}$, which does not affect the $\arg\max$, and
perturbed by only $O(\alpha/\sqrt{D})$.

\emph{Balanced class signs.} Prototypes of different classes typically share a large common component (e.g., over $90\%$ of the prototype energy in Vanilla HDC, as implied by the measurements in Section~\ref{sec:dimension}). If $w$ were added with the same sign to every class, this component would act as strong interference when reading the watermark. Balanced signs let verification cancel it exactly (Section~\ref{sec:analysis}).

\subsection{Verification}
Given the prototypes $\tilde{P}$ of a suspect model, the verifier
regenerates $b$, $R_i$, and $s$ from $(K, I)$; neither the original
model nor any data is needed. It then computes
(Fig.~\ref{fig:method}, \ding{184})
\begin{equation}
e = \frac{1}{C} \sum_{c=1}^{C} s_c \frac{\tilde{P}_c}{\lVert \tilde{P}_c \rVert},
\qquad
\hat{b}_i = \mathbf{1}\big[\langle R_i, e \rangle > 0\big].
\label{eq:verify}
\end{equation}
Normalizing each prototype removes any rescaling and gives every class an equal vote. Let $m$ be the number of positions where $\hat{b}$ and $b$ agree. If the suspect model is unrelated to the key, the carriers are independent of $e$, so each bit agrees with probability $1/2$ regardless of the model. The evidence is therefore the one-sided binomial $p$-value
\begin{equation}
p = \sum_{k=m}^{L} \binom{L}{k} 2^{-L},
\label{eq:pvalue}
\end{equation}
and ownership is claimed at significance level $\tau$ if $p < \tau$.

\subsection{Analysis}
\label{sec:analysis}

Let $u_c = P_c/\lVert P_c\rVert$ be the normalized clean prototypes, $\mu = \frac{1}{C}\sum_c u_c$ their shared component, and $\Delta_c = u_c - \mu$ the class-specific remainders. Since $w$ is quasi-orthogonal to $P_c$, normalizing Eq.~\eqref{eq:embed} gives $u^{\mathrm{wm}}_c \approx (u_c + \alpha s_c w)/\sqrt{1+\alpha^2}$, and Eq.~\eqref{eq:verify} becomes

\begin{equation}
e \approx \frac{e_0 + \alpha\, w}{\sqrt{1+\alpha^2}}, \qquad
e_0 = \frac{1}{C}\sum_{c} s_c u_c = \frac{1}{C}\sum_{c} s_c \Delta_c .
\label{eq:e}
\end{equation} 

The watermark adds coherently over classes because $s_c^2 = 1$, while $\mu$ vanishes from the host term $e_0$ because $\sum_c s_c = 0$. Projecting $e$ onto a carrier, $\langle R_i, w\rangle \approx \bar{b}_i/\sqrt{L}$ by Eq.~\eqref{eq:w}, whereas $\langle R_i, e_0\rangle$ is zero-mean noise with standard deviation $\lVert e_0\rVert/\sqrt{D}$, since $R_i$ is independent of the host model. Each bit is thus recovered with probability

\begin{equation}
\Pr[\hat{b}_i = b_i] \approx
\Phi\!\left(\frac{\alpha\sqrt{D}}{\sqrt{L}\,\lVert e_0\rVert}\right)
\approx
\Phi\!\left(\alpha\sqrt{\frac{(C-1)\,D}{L\,(1-\lVert\mu\rVert^2)}}\right),
\label{eq:snr}
\end{equation}

where $\Phi$ is the standard normal distribution function and the second form uses $\mathbb{E}_s\lVert e_0\rVert^2 = (1-\lVert\mu\rVert^2)/(C-1)$ for random balanced signs. Eq.~\eqref{eq:snr} predicts that bit recovery improves with the dimension $D$ and trades off against the payload length $L$; we verify this prediction in Section~\ref{sec:dimension}. Finally, because the representation is holographic, the watermark is spread evenly over all $D$ components of every prototype, so an attack that damages part of the prototypes removes only a corresponding part of the watermark.

\section{Evaluation}
\label{sec:eval}

\subsection{Experimental Setup}

\textbf{Datasets and models.} We evaluate HDMark on MNIST~\cite{lecun1998gradient}, FashionMNIST~\cite{xiao2017fashion},
ISOLET~\cite{fanty1990spoken}, and UCI-HAR~\cite{anguita2013public} across the three learning rules of Section~\ref{sec:prelim} with $D = 10{,}000$: Vanilla HDC (single pass) and AdaptHD
(30 epochs) use record-based encoding, and OnlineHD (5 epochs) uses
nonlinear random projection~\cite{heddes2023torchhd}.

\textbf{Protocol.} The official training set of each dataset is split
into $80\%$ for the owner's training, $10\%$ for validation, and $10\%$ held
out for the adversary; all results are reported on the official test set.
The payload has $L = 64$ bits. For each dataset and model, $\alpha$ is
the smallest value on the validation grid that verifies at $p < 10^{-6}$
in at least $95\%$ of validation trials while keeping the accuracy drop
below 0.5\,percentage points (pp) and prediction agreement with the
clean model above $99\%$ (Table~\ref{tab:main}). Every configuration is
repeated over 5 training seeds and 10 independent keys per seed
(50 trials).

\begin{table*}[t]
\centering
\caption{Fidelity and verification of HDMark at the calibrated $\alpha$ ($L{=}64$, $D{=}10{,}000$; 5 seeds $\times$ 10 keys per row), compared with a trigger-set black-box watermark trained on the same split (5 seeds per row). $\Delta$Acc: accuracy change in percentage points; Agr.: prediction agreement with the clean model; VR: verification rate.}
\label{tab:main}
\footnotesize
\begin{tabular}{ll c c ccccc ccc}
\toprule
 & & & & \multicolumn{5}{c}{\textbf{HDMark}} & \multicolumn{3}{c}{\textbf{Black-box (trigger set)}} \\
\cmidrule(lr){5-9}\cmidrule(lr){10-12}
Dataset & Model & $\alpha$ & Clean Acc & $\Delta$Acc & Agr. & BRR & VR$_{10^{-3}}$ & VR$_{10^{-6}}$ & $\Delta$Acc & Agr. & VR$_{10^{-3}}$ \\
\midrule
\multirow{3}{*}{MNIST} & Vanilla & 0.01 & 82.66 & 0.00 & 99.9 & 97.4 & 100.0 & 100.0 & -0.62 & 95.4 & 60.0 \\
 & AdaptHD & 0.04 & 94.96 & -0.02 & 99.8 & 93.0 & 100.0 & 100.0 & -0.47 & 96.6 & 100.0 \\
 & OnlineHD & 0.015 & 96.65 & -0.02 & 99.9 & 86.2 & 100.0 & 90.0 & 0.00 & 97.0 & 100.0 \\
\midrule
\multirow{3}{*}{FashionMNIST} & Vanilla & 0.01 & 69.53 & 0.00 & 99.9 & 90.8 & 98.0 & 94.0 & -4.18 & 75.5 & 20.0 \\
 & AdaptHD & 0.03 & 84.01 & -0.01 & 99.7 & 87.3 & 100.0 & 96.0 & -0.81 & 88.8 & 100.0 \\
 & OnlineHD & 0.015 & 84.87 & 0.00 & 99.6 & 88.3 & 100.0 & 98.0 & -0.40 & 85.8 & 100.0 \\
\midrule
\multirow{3}{*}{ISOLET} & Vanilla & 0.01 & 88.49 & +0.01 & 99.9 & 98.7 & 100.0 & 100.0 & -0.63 & 97.4 & 100.0 \\
 & AdaptHD & 0.02 & 93.41 & 0.00 & 99.9 & 89.3 & 100.0 & 100.0 & +0.09 & 95.9 & 100.0 \\
 & OnlineHD & 0.015 & 93.44 & -0.01 & 100.0 & 89.7 & 100.0 & 100.0 & -0.26 & 93.8 & 100.0 \\
\midrule
\multirow{3}{*}{UCI-HAR} & Vanilla & 0.02 & 84.47 & +0.05 & 99.7 & 97.7 & 100.0 & 98.0 & -8.29 & 81.4 & 40.0 \\
 & AdaptHD & 0.05 & 93.08 & +0.01 & 99.8 & 92.2 & 100.0 & 98.0 & -0.24 & 95.9 & 100.0 \\
 & OnlineHD & 0.04 & 92.11 & +0.03 & 99.6 & 96.7 & 100.0 & 94.0 & +0.37 & 94.8 & 100.0 \\
\midrule
\multirow{3}{*}{Average} & Vanilla & -- & 81.29 & +0.01 & 99.9 & 96.1 & 99.5 & 98.0 & -3.43 & 87.4 & 55.0 \\
 & AdaptHD & -- & 91.37 & 0.00 & 99.8 & 90.5 & 100.0 & 98.5 & -0.36 & 94.3 & 100.0 \\
 & OnlineHD & -- & 91.77 & 0.00 & 99.8 & 90.2 & 100.0 & 95.5 & -0.07 & 92.9 & 100.0 \\
\bottomrule
\end{tabular}
\end{table*}


\textbf{Metrics.} We report the accuracy change $\Delta$Acc, the
prediction agreement with the clean model, the bit recovery rate (BRR,
the fraction of payload bits recovered), and the verification rate (VR,
the fraction of trials with $p < \tau$). For $L = 64$, the significance
levels $\tau = 10^{-2}$, $10^{-3}$, and $10^{-6}$ require at least 42,
45, and 51 matching bits, i.e., a BRR of $65.6\%$, $70.3\%$, and $79.7\%$. We
use $\tau = 10^{-3}$ unless stated otherwise.

\subsection{Fidelity and Verification}

Table~\ref{tab:main} shows that HDMark preserves task accuracy and prediction agreement: accuracy changes by at most 0.05\,pp and at least $99.6\%$ of the
predictions are identical to those of the clean model, in line with
Eq.~\eqref{eq:fidelity}. At the same time, $86-99\%$ of the payload bits are recovered, far above
the $70.3\%$ required. The VR is therefore $100\%$ in 11 of the 12
configurations and $98\%$ in the remaining one (Vanilla HDC on
FashionMNIST); even at $\tau = 10^{-6}$ it is $90-100\%$.

\textbf{False positives.} A key that is unrelated to the model should
match each bit with probability $1/2$, in which case Eq.~\eqref{eq:pvalue}
gives a probability of $0.84\%$ and $0.078\%$ of reaching the 42- and 45-bit
thresholds of $\tau = 10^{-2}$ and $10^{-3}$. We test 60{,}000 wrong
keys on the watermarked models: they recover $50.0\%$ of the bits and are
accepted at rates of $0.80\%$ and $0.07\%$, as predicted, and never at
$\tau = 10^{-6}$. A further 3{,}600 trials with the correct key on unmarked, independently
trained, and randomly perturbed models give the same picture, with only $0.06\%$
accepted at $\tau = 10^{-3}$ and none at $\tau = 10^{-6}$.

\textbf{Post-training vs.\ in-training watermarking.}
Table~\ref{tab:main} also compares HDMark against the black-box
trigger-set watermark of~\cite{poursiami2026intellectual} on the same
data splits. Trigger sets enable query-only verification, but they must
be learned jointly with the task, which shifts the prototypes toward
out-of-distribution trigger samples. In single-pass Vanilla HDC,
superposition dilution leaves trigger verification unreliable
($20-60\%$ VR on three datasets), while the embedding still costs up to
$8.3$\,pp of accuracy and alters up to $24.5\%$ of the test predictions.
Even with the iterative updates of AdaptHD and OnlineHD, fitting the
triggers changes $3-14\%$ of the predictions. HDMark instead superimposes
a hypervector that is quasi-orthogonal to the trained prototypes. It
therefore requires no retraining, preserves at least $99.6\%$ of the
predictions under all three learning rules, and lets the owner stamp a
distinct $L$-bit identity into each distributed copy of a single trained
model.

\subsection{Robustness to Removal Attacks}
\label{sec:attacks}

\begin{figure*}[t]
  \centering
  \includegraphics[width=\linewidth]{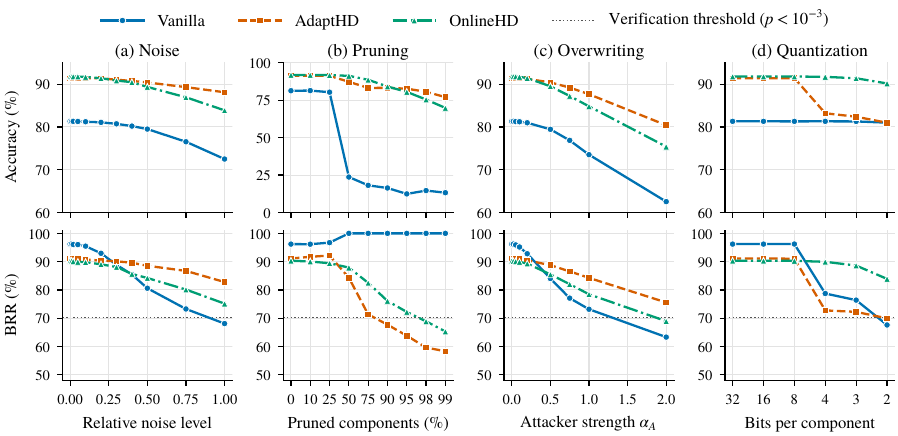}
  \caption{Accuracy (top) and bit recovery rate (bottom) of the
  watermarked models under (a) noise, (b) pruning, (c) overwriting, and
  (d) quantization, pooled over the four datasets. The dotted line is
  the BRR required for verification at $\tau = 10^{-3}$ (45 of 64
  bits).}
  \label{fig:attacks}
\end{figure*}

We consider five attacks on the prototypes; Fig.~\ref{fig:attacks} and
Table~\ref{tab:perm} report results pooled over the datasets (100
trials per point).

\textbf{Fine-tuning.} The adversary continues training the stolen model
with its native learning rule on their own data ($10\%$ of the training
set). Verification remains at 100\% for all three models and the BRR
drops by at most $1.2$\,pp (Table~\ref{tab:perm}). This is expected:
fine-tuning adds encoded samples to the prototypes, and these are
quasi-orthogonal to the secret carriers, so the updates dilute the
watermark slightly but cannot cancel it.

\textbf{Noise.} The adversary adds Gaussian noise whose norm is a
fraction $r$ of each prototype's norm (Fig.~\ref{fig:attacks}(a)). The BRR
decreases smoothly with $r$. It stays above the $70.3\%$ threshold for
AdaptHD and OnlineHD even at $r = 1$, where the noise is as large as the
prototypes themselves, and falls below it only for Vanilla HDC at
$r = 1$, at a cost of $8.8$\,pp of accuracy.

\textbf{Pruning.} The adversary zeroes the smallest-magnitude components
of each prototype (Fig.~\ref{fig:attacks}(b)). For AdaptHD and OnlineHD,
the BRR declines gradually with the pruned fraction, as expected from a
holographic watermark, and crosses the threshold only at $90\%$ and $98\%$
pruning, where the models have lost $8.2$ and $16.4$\,pp of accuracy. Vanilla HDC loses its accuracy at $50\%$ pruning while its watermark remains fully readable. Under record-based encoding, roughly half of the dimensions stay constant across inputs; because Vanilla HDC bundles all samples unconditionally, these uninformative dimensions add coherently into the largest prototype entries. Magnitude pruning therefore discards the smaller, class-discriminative entries first, leaving a class-common background that the balanced signs $s_c$ cancel during verification, so the watermark is read without interference.

\textbf{Overwriting.} The adversary embeds a second HDMark watermark
with their own key and strength $\alpha_A$ (Fig.~\ref{fig:attacks}(c)).
Since the two sets of carriers are quasi-orthogonal, the second
watermark acts on the first only as noise. The BRR stays above the
threshold up to $\alpha_A = 1$, at least 20 times the owner's strength,
and falls below it for Vanilla HDC and OnlineHD only at $\alpha_A = 2$,
at a cost of $18.7$ and $16.4$\,pp. The adversary's own watermark does not
create a competing claim, because the owner's timestamped commitment predates it.

\textbf{Quantization.} The adversary quantizes each prototype to $q$ bits
using a symmetric uniform quantizer with MSE-optimal clipping to best
preserve the stolen model (Fig.~\ref{fig:attacks}(d)). Down to $8$ bits,
both accuracy and BRR are unchanged ($100\%$ VR for all models). Under
coarser quantization ($4$ to $2$ bits), OnlineHD stays well above the
threshold ($99\%$ VR at $2$ bits). AdaptHD approaches
the threshold ($72.8\%$ to $69.9\%$ BRR), but at these precisions the
quantized model is itself $8.2$ to $10.4$\,pp less accurate, so its
watermark weakens only as the model loses its utility.
In Vanilla HDC (which is $\sim10$\,pp less accurate overall), the model keeps
its accuracy down to $2$ bits, but because it is calibrated with the
smallest $\alpha$, the watermark offset amounts to less than $8\%$ of a
quantization step and the BRR dips below the threshold at $2$ bits
($67.6\%$). Even there, because HDMark is embedded in a single
post-training step with ample fidelity headroom, the owner can simply
choose a wider margin when targeting low-precision deployment: raising
Vanilla HDC's mean $\alpha$ from $0.0125$ to $0.0375$ changes
accuracy by at most $0.04$\,pp while lifting
verification across $4$ to $2$ bits to at least $97\%$.

Overall, no removal attack brings the mean BRR below the $\tau = 10^{-3}$
threshold without costing over $8$\,pp of accuracy, except under $2$-bit
quantization of Vanilla HDC, where a modestly larger embedding margin
restores verification at negligible accuracy cost.

\subsection{Permutation Attack and Realignment}
\label{sec:permutation}
Because cosine similarity is invariant to a common reordering and sign
change of the coordinates, an adversary can apply the same secret
permutation $\pi$ and sign flips $\sigma \in \{\pm 1\}^{D}$ to the
encoder output and to every prototype. This preserves every prediction
exactly, yet misaligns the prototypes with the owner's carriers $R_i$
and reduces the BRR to chance (Table~\ref{tab:perm}).

\begin{table}[hb]
\centering
\caption{The effect of \textbf{Fine-tuning} (10\% of the training data) and hypervector
\textbf{Permutation} with sign flips, without and with encoder
\textbf{Realignment}. Datasets pooled; VR at $\tau = 10^{-3}$.}
\label{tab:perm}
\footnotesize\setlength{\tabcolsep}{2.1pt}
\begin{tabular}{ccc ccc ccc ccc}
\toprule
 & & & \multicolumn{3}{c}{\textbf{Vanilla}} & \multicolumn{3}{c}{\textbf{AdaptHD}} & \multicolumn{3}{c}{\textbf{OnlineHD}} \\
\cmidrule(lr){4-6}\cmidrule(lr){7-9}\cmidrule(lr){10-12}
Fine-tuning & Permutation & Realignment& & BRR & VR & & BRR & VR & & BRR & VR \\
\midrule
-- & -- & -- & & 96.2 & 100 & & 91.1 & 100 & & 90.3 & 100 \\
\checkmark & -- & -- & & 95.0 & 100 & & 90.6 & 100 & & 89.8 & 100 \\
\midrule
-- & \checkmark & -- & & 50.4 & 0 & & 48.7 & 0 & & 49.4 & 0 \\
-- & \checkmark & \checkmark & & 84.1 & 95 & & 74.3 & 79 & & 90.3 & 100 \\
\checkmark & \checkmark & -- & & 50.4 & 0 & & 49.0 & 0 & & 49.6 & 0 \\
\checkmark & \checkmark & \checkmark & & 80.9 & 90 & & 73.5 & 72 & & 89.4 & 100 \\
\bottomrule
\end{tabular}
\end{table}

To keep inference intact, the adversary must apply the same
$(\pi, \sigma)$ to the encoder, so the suspect encoder reveals the
transformation. We exploit this by \emph{realignment}. The verifier
feeds $M$ probe inputs through the original and suspect encoders to
obtain encoding matrices $H, \tilde{H} \in \mathbb{R}^{M \times D}$, and
matches each column of $\tilde{H}$ to the column of $H$ with which it
has the highest absolute correlation; the sign of that correlation gives
the flip. Applying the recovered inverse map to the suspect prototypes
before verification fully neutralizes the attack for OnlineHD, whose
nonlinear projection varies with the input in every coordinate: the VR
returns to $100\%$, even when the attack is combined with fine-tuning.

Under record-based encoding, the input-independent dimensions noted in
Section~\ref{sec:eval}-C take the same value for every probe input, so their columns in $H$ are indistinguishable from one another and cannot be aligned, whereas the input-varying dimensions align cleanly. Because HDMark is holographic, this aligned subset alone recovers $74-84\%$ of the payload bits and restores verification in $95\%$ (Vanilla HDC) and $79\%$ (AdaptHD) of trials.

\begin{figure}[b!]
  \centering
  \includegraphics[width=\linewidth]{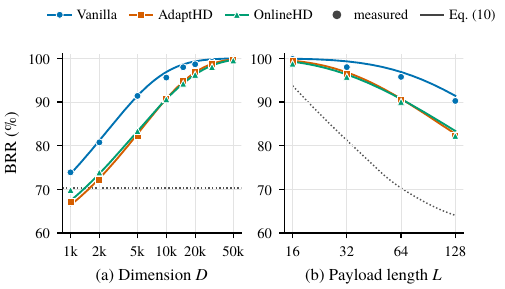}
  \caption{Bit recovery rate versus (a) dimension $D$ and (b) payload
  length $L$ at the calibrated $\alpha$, pooled over the datasets.
  Markers are measurements; lines are the prediction of
  Eq.~\eqref{eq:snr} with $\lVert e_0\rVert$ computed from the unwatermarked prototypes. The dotted line is the verification threshold at $\tau = 10^{-3}$ Eq.~\eqref{eq:pvalue}.}
  \label{fig:dim}
\end{figure}

\begin{figure*}[t!]
  \centering
  \includegraphics[width=\linewidth]{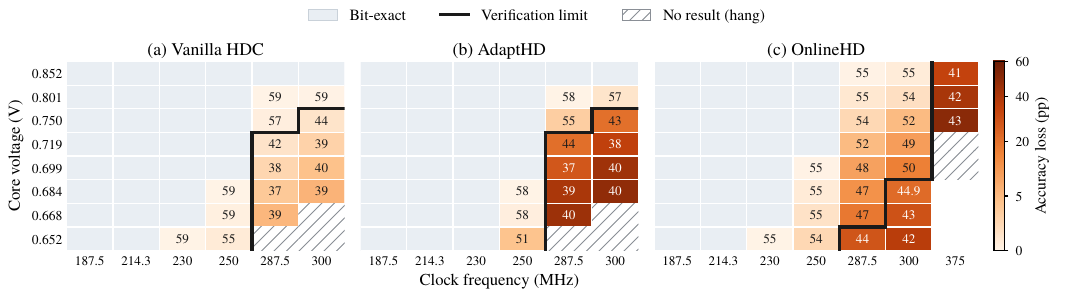}
  \caption{Voltage and frequency scaling on the FPGA, pooled over the four datasets. Color: accuracy loss relative to the nominal run. Numbers inside the cells: payload bits recovered on-chip (of 64). The thick line marks the verification limit ($45$ bits, $\tau = 10^{-3}$).}
  \label{fig:fpga}
\end{figure*}

\subsection{Effect of Dimension and Payload Length}
\label{sec:dimension}
We now test the scaling with $D$ and $L$ predicted by Eq.~\eqref{eq:snr}.
When the models are retrained across $D \in [10^3, 5\times 10^4]$ with $L = 64$
and $\alpha$ unchanged (Fig.~\ref{fig:dim}(a)), the BRR rises
from $67-74\%$ at $D = 1{,}000$ to above $99.6\%$ at $D = 50{,}000$, and
verification reaches $100\%$ from $D = 10{,}000$ onward.  Likewise, at fixed $D = 10{,}000$ and $\alpha$, the BRR decreases from $99-100\%$ at $L = 16$ to $82-90\%$ at $L = 128$ (Fig.~\ref{fig:dim}(b)). A longer payload still yields stronger statistical evidence, because more bits are tested in~\eqref{eq:pvalue}: at $L = 16$, even perfect recovery gives $p = 2^{-16} \approx 1.5 \times 10^{-5}$ and can never pass $\tau = 10^{-6}$, whereas $L = 64$ and $L = 128$ verify at $\tau = 10^{-6}$ in $94-98\%$ of trials. 

Beyond these trends, Eq.~\eqref{eq:snr} predicts the BRR quantitatively (lines in Fig.~\ref{fig:dim}):
with $\lVert e_0\rVert$ computed from the unwatermarked prototypes, it
matches the measured BRR with a mean absolute error of 0.4\,pp over 288
settings (14{,}400 trials) varying $D$, $L$, and $\alpha$ across all
models and datasets. The strength $\alpha$ needed for a target reliability can thus be
determined in closed form from the trained prototypes, without an
empirical sweep.

\section{Hardware Evaluation}
\label{sec:hardware}

\textbf{Implementation.} We implement all three classifiers on an AMD
ZCU104 board (Zynq UltraScale+ MPSoC). The programmable logic runs the
complete pipeline at $D = 10{,}000$: the encoder, the similarity search
over the prototypes, which are stored as 16-bit integers in on-chip
memory, and the watermark verification. Verification needs no dedicated
logic: by Eq.~\eqref{eq:verify}, $\langle R_i, e\rangle$ is a signed sum of
the similarities between the carrier $R_i$ and the prototypes, so each
carrier is scored like an encoded input by the same datapath. The nominal operating point is a $0.852$\,V core supply and a $187.5$\,MHz clock.\\
\indent \textbf{Overhead.} At the nominal point, all models reproduce the software reference bit for bit, including every recovered payload bit. HDMark changes only the contents of the prototype memory, so the circuit is unchanged, and the measured inference energy with and without the watermark differs by less than $0.4\%$, within the resolution of the power sensor. One inference takes $28-36\,\mu$s and $76-103\,\mu$J on the core supply for Vanilla HDC and AdaptHD ($92-133\,\mu$s and $295-459\,\mu$J for OnlineHD).

\textbf{Reliability under voltage and frequency scaling.} We lower the core voltage from $0.852$ to $0.652$\,V and raise the clock from $187.5$ up to $300$\,MHz ($375$\,MHz for OnlineHD) until the hardware hangs, comparing every output with the nominal run (Fig.~\ref{fig:fpga}). At the nominal clock, all designs stay bit-exact down to $0.652$\,V, the lowest voltage we tested. At this level, the core energy per inference falls by $45-46\%$ while every prediction and every payload bit is unchanged. HDMark thus remains completely reliable under aggressive low-power voltage scaling.

Combining overclocking with reduced voltage eventually induces timing errors. For AdaptHD and OnlineHD, on-chip verification holds until these faults severely degrade the classifier itself. Specifically, every operating point with up to $18$\,pp accuracy loss remains verified ($\ge 45$ bits), and at the points where verification is lost, accuracy has dropped by $34.7$\,pp on average for AdaptHD and $36.9$\,pp for OnlineHD (at least $20.1$ and $21.4$\,pp, respectively).

Vanilla HDC carries the smallest $\alpha$ (Sec.~\ref{sec:attacks}), and this narrower embedding margin makes its on-chip check more sensitive to timing errors. For example, at $287.5-300$\,MHz, verification is lost at an average accuracy loss of $2.8$\,pp. This drop comes from errors in the on-chip arithmetic and not from a loss of the watermark. When the prototypes are read out of the board at the same stressed settings and verified off-chip, ownership is established in all runs, although up to $7.9\%$ of the bits read out are wrong; the memory contents themselves are unchanged (bit-error rate below $10^{-5}$ once the stress is removed).

Finally, timing faults cause no false ownership claims, as neither an unmarked model nor a wrong key passes verification at any operating point.

\section{Conclusion}
\label{sec:conclusion}

We presented HDMark, the first native, multi-bit model watermarking framework built on the algebra of hyperdimensional computing. By superimposing quasi-orthogonal carriers with balanced class signs onto the trained class prototypes, HDMark embeds a verifiable owner identity without retraining and preserves clean predictions across both single-pass and iterative HDC learners. Moreover, the geometry of this superposition yields closed-form expressions for fidelity and bit recovery that match empirical measurements within $0.4$\,pp mean absolute error. Across four benchmarks, the watermark survives fine-tuning and low-bit quantization, resists the evaluated attacks under the reported settings, and recovers from HDC-specific dimension permutations through encoder
realignment. Finally, our AMD ZCU104 FPGA implementation confirms that HDMark reuses the inference datapath for verification with no hardware overhead and remains reliable under aggressive voltage scaling and timing faults, enabling practical IP protection for edge HDC accelerators.

\clearpage
\bibliographystyle{IEEEtran}
\bibliography{ref}
\end{document}